\documentstyle[11pt,newpasp,epsf]{article}

\begin{document}

\title{Preliminary analysis of an extreme helium sdO star:~ BD+25 4655}

\author{J. Budaj}
\affil{Astronomical Institute
05960 Tatransk\'{a} Lomnica, Slovak Republic, budaj@ta3.sk}

\author{V. Elkin\altaffilmark{1}}
\affil{SAO, Nizhnij Arkhyz, Russia, vgelk@sao.ru}

\author{I. Hubeny}
\affil{National Optical Astronomy Observatories, Tucson, USA, 
hubeny@tlusty.gsfc.nasa.gov}



\altaffiltext{1}{Visiting Astronomer, 
University of Central Lancashire, UK}
\altaffiltext{2}{Note: this is a poster from the IAU Symp. No.210.
Posters were distributed on a CD-rom and as such, instead of the page
numbers, it has a cod E44 assigned to it.}


\setcounter{footnote}{1}


\begin{abstract}

Preliminary analysis of CCD spectra obtained  by the 6m SAO telescope is
presented.  We have used simple H-He NLTE  model atmospheres  computed by
TLUSTY to derive the basic parameters of the star.

\end{abstract}




%
%
%

\section{Introduction}
\label{s1}

BD+25 4655 (SAO 90153, HIP108578, IS Peg) is an interesting object known 
today as a variable O subdwarf, suspected binary and spectrophotometric 
standard at the same time.
It was mentioned as an sdO star in Greenstein (1960) already.
First detail analysis of this star using LTE approach was accomplished
by Peterson (1970). He derived the following basic parameters:
$T_{eff}=43000 K, \log g=6.7, Y/X=49, \log L/L_{\odot}=1.3, M/M_{\odot}=1.2$ 
(see also Richter 1971).
Greenstein \& Sargent (1974) list equivalent widths of some lines
and $T_{eff}=42000 K, \log g=6.7, M_{V}=6.0$.
Some spectroscopic data were published also in an atlas of spectra of He-rich
stars of Kaufmann \& Theil (1980) covering 3700-4600\AA\AA.
Bartolini et al. (1982) searched for variability in hydrogen poor 
stars and although they failed to find regular periodicities for this star
they did find variations in UBV with a period of
$P=0.009368^{d}$ and amplitude of $\Delta m=0.03^{m}$ in two nights. 
At the same time, 
they suggested long term variations on a time scale of several months 
and amplitude of $\Delta V=0.07^{m}$.
Dworetsky et al. (1982) measured the star in UBV system and obtained:
V=9.69, B-V=-0.26, U-B=-1.16, E(B-V)=0.06. They classified it as O4:.
Later on Colina \& Bohlin (1994) obtained V=9.656, B-V=-0.305.
Tobin (1985) obtained uvby$\beta$ photometry.
Bartkevicius \& Lazauskaite (1996) classified the star in the Vilnius 
photometric system as: sdO6, He, and estimated $E(Y-V)=0, M_{V}=4.0$.
Diplas \& Savage (1994) set an upper limit for interstellar
HI column number density in front of the star: $log N(HI)=19.94$ 
per $cm^{-2}$. 
The star is a HST optical and UV spectrophotometric standard and
Oke (1990) measured its absolute spectral energy distribution in the range
3200-9200 \AA\AA~ while Bohlin et al. (1990) obtained absolute UV flux 
from IUE data.
HIPPARCOS gives the parallax $\pi=8.99 \pm 1.20$ mas (ESA 1997)
and labels the star as constant. Above parallax corresponds to the distance 
of 111 pc.
Elkin (1998) attempted to measure the magnetic field but could set only
an upper limit of about 300 G and concluded that there is no field within 
the precision of his measurements. 
Ulla \& Thejll (1998) did IR photometry and obtained:
$J=10.39, H=10.52, K=10.58$. They found an excess in JHK fluxes
of about $\Delta J=0.1^{m}$ and interpreted it 
as due to the binary companion of the sp. type earlier than B8.
They also put the following upper limit for the reddening
$E(B-V)<0.025^{m}$.

Main aim of the paper is to present a preliminary analysis of this star 
based on high resolution, high signal to noise spectra using pure 
NLTE H-He model atmospheres as well as to  compile other available data 
from the literature for a more elaborated study.
Similar NLTE analysis of sdO stars were recently 
carried by Lanz et al. (1997) (using line blanketed NLTE models) 
and Thejll et al. (1994).
Recent review on sdO stars can be found in Heber (1992).

\section{Observations}

The high resolution spectra were obtained at 6-m telescope
of SAO RAS using Main stellar spectrograph located
in Nasmith 2 platform of the telescope.
Spectral resolution of spectra obtained in blue and red regions are 0.3 \AA
~and 0.45 \AA, respectively.
For both observations and subsequent data reduction we used MIDAS package
and also DECH20 code (Galazutginov 1992).
The low resolution spectra were obtained at 1-m telescope with 
UAGS spectrograph. Spectral resolution of this spectrum is about 6 \AA.
Here the data reduction were done using the software written
by Vlasyuk (1993) and DECH20 of Galazutginov (1992).

\begin{table}
\caption{Log of observations: 
spectrum identifier, date [ddmmyyyy], time [hhmm], exposure [m],
JD-2400000 of the middle of the exposure, wavelength interval covered
[\AA\AA], resolution [\AA], heliocentric correction [km/s] 
and radial velocity [km/s]. 
Note: the second spectrum is not suitable for RV measurements and first
spectrum has rather large error.}
\begin{center}
\begin{tabular}{lllllllll}
\hline
Sp.  &date      &  UT  &E &JD        &$\lambda$ &R & h.c.&RV \\
\hline
1 &13 07 1997&21 54&30&50643.423 &6489-6732&0.45 &+18.1&-37.2$\pm 2.3$ \\
2 &15 09 1997&20 05&30&50707.347 &4352-4514&0.3  & --  &  --           \\
3 &18 09 1997&18 12&30&50710.269 &4352-4514&0.3  &-5.6 &-30.6$\pm 0.3$ \\
4 &07 09 1998&19 38&30&51064.328 &4320-4482&0.3  &-1.1 &-31.7$\pm 1.0$ \\
5 &08 06 2001&01 43& 5&52069.406 &3738-5441&6    &     &               \\
\hline
\end{tabular}
\end{center}
\label{t1}
\end{table}

\section{NLTE calculations}

For the calculation of NLTE atmosphere models and level populations 
of explicit ions we used the {\sc {tlusty195}} code described in more 
detail in 
Hubeny (1988), Hubeny \& Lanz (1992) and Hubeny \& Lanz (1995). 
Table \ref{t2} lists which ions and how many levels were 
treated explicitly
what means that their level populations were calculated in NLTE and
their opacity was considered. Model of atoms were constructed 
using and IDL interface tool MODION developed by 
Varosi et al.(1995) from the Opacity Project Data. 
Other elements up to $Z=30$ were
allowed to contribute to the particle and electron number density in LTE.
Synthetic spectra were calculated with the {\sc{synspec42}} code 
Hubeny et al. (1995). Atomic data for line transitions were taken from
Kurucz (1990) but VALD data base was also consulted (Kupka et al. 1999).

\begin{table}
\caption{Number of explicit levels considered in particular explicit 
elements/ions; abundances estimated.}
\begin{center}
\begin{tabular}{crlllll}
\hline
element/ion  &  I &  II & III  &  IV & V &  A \\
\hline
H            &  9 &     &      &     &   &  1 \\
He           & 14 & 14  &      &     &   & 40 \\
\hline
\end{tabular}
\end{center}
\label{t2}
\end{table}

\section{Discussion}

\begin{figure}
\plotone{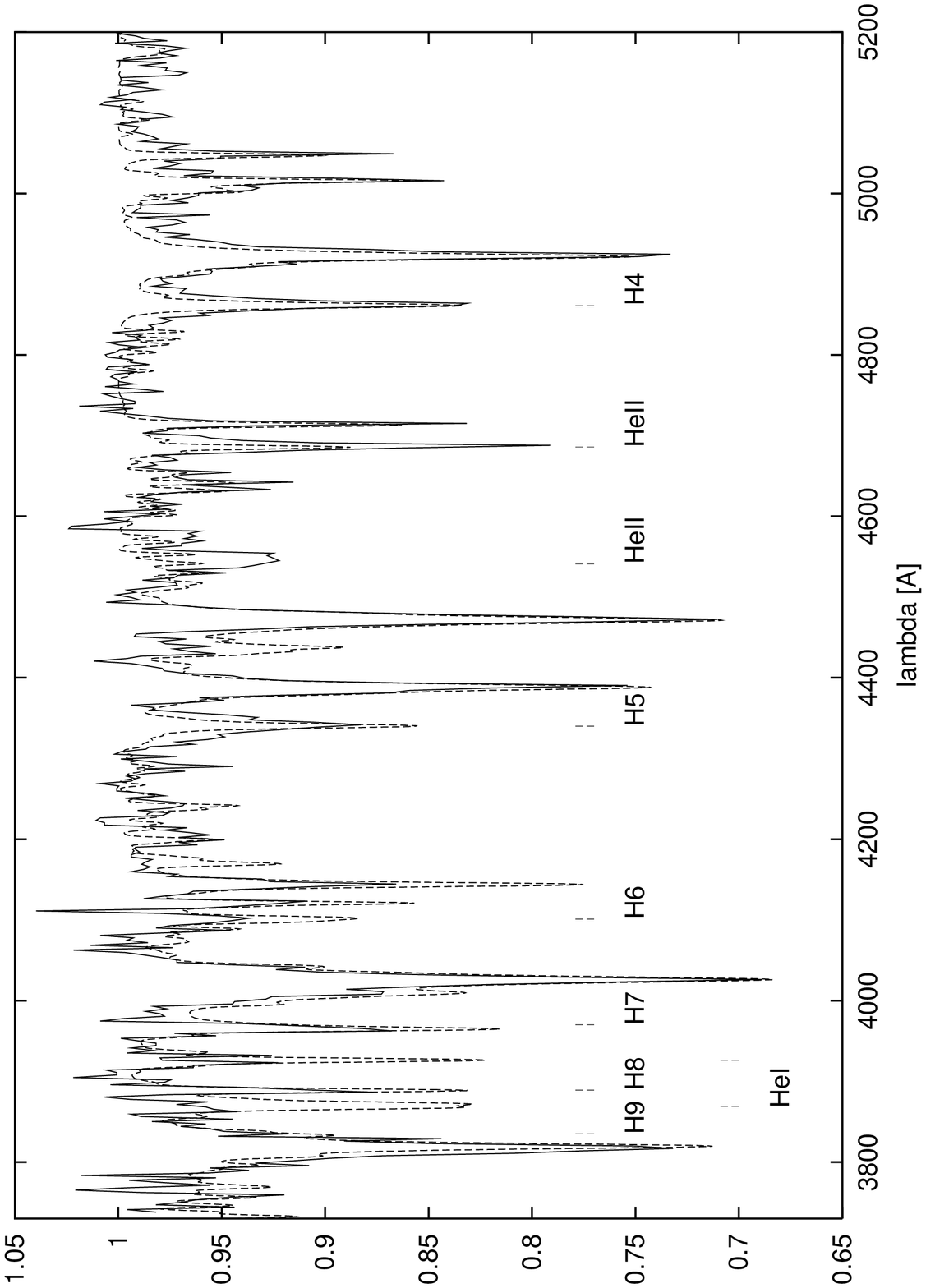}
\caption{Relative intensity of the low resolution spectrum of BD+254655.
Solid line - observations, dashes -synthetic spectrum for
$T_{eff}=38000; \log g=5.3; N(He/H)=40$.}
\label{f1}
\end{figure}

\begin{figure}
\plotone{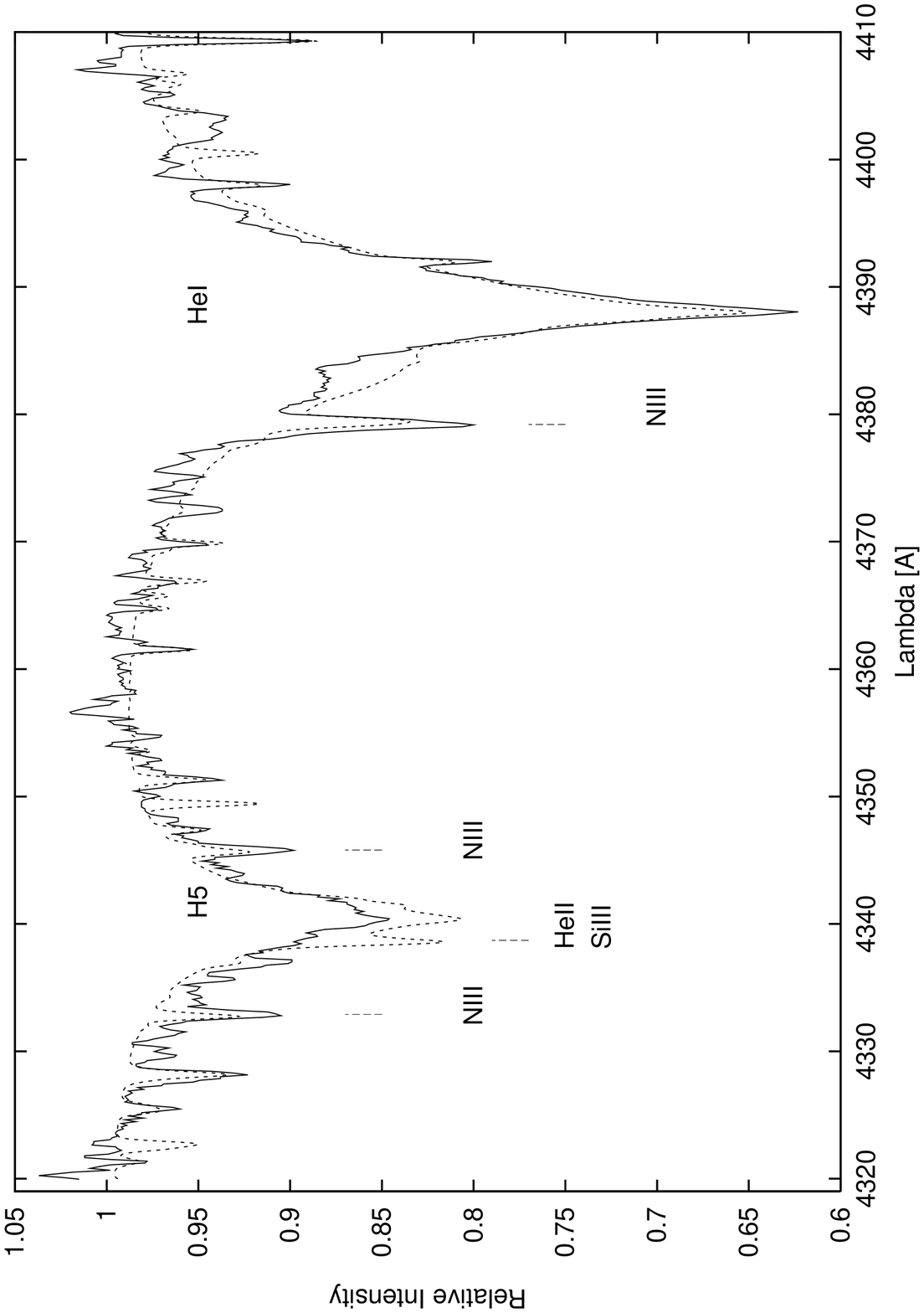}
\caption{High resolution spectrum of BD+254655.
Solid line - observations, dashes -synthetic spectrum for  
$T_{eff}=38000; \log g=5.3; N(He/H)=40$.}
\label{f2}
\end{figure}

\begin{figure}
\plotone{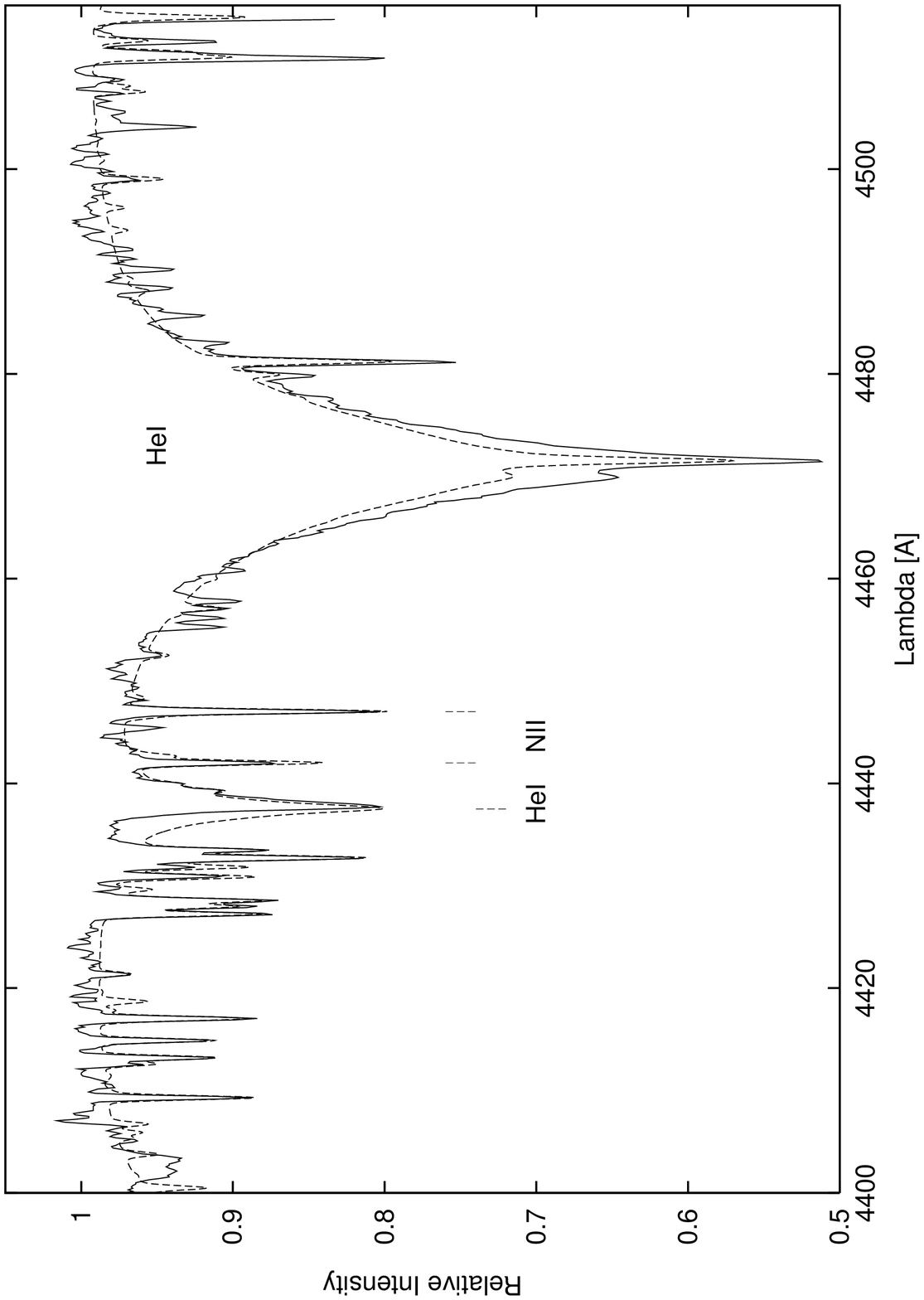}
\caption{High resolution spectrum of BD+254655.
Solid line - observations, dashes -synthetic spectrum for  
$T_{eff}=38000; \log g=5.3; N(He/H)=40$.}
\label{f3}
\end{figure}

\begin{figure}
\plotone{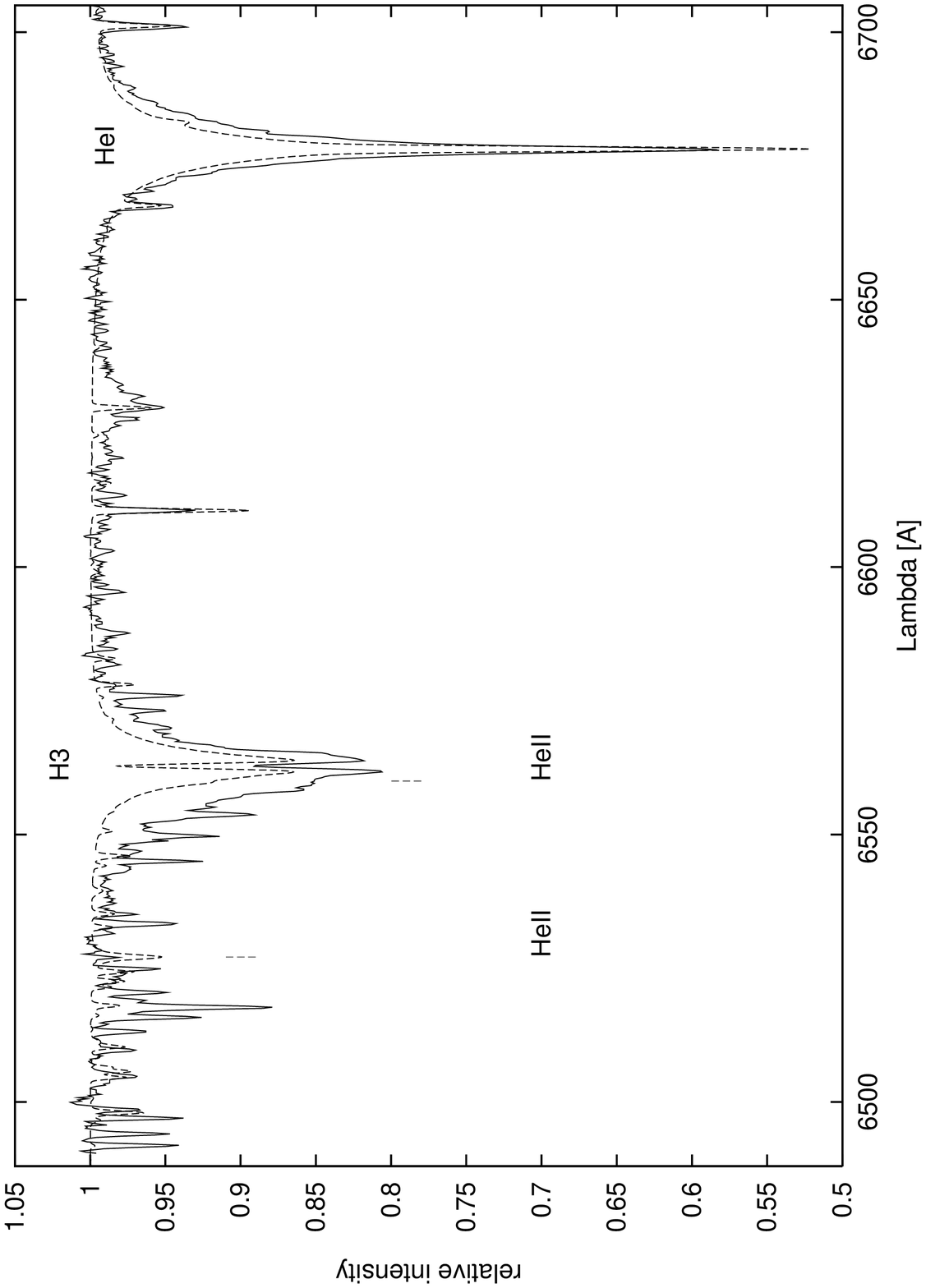}
\caption{High resolution spectrum of BD+254655.
Solid line - observations, dashes -synthetic spectrum for  
$T_{eff}=38000; \log g=5.3; N(He/H)=40$.}
\label{f4}
\end{figure}

The spectrum of this star is hard to fit and understand. 
We will describe various interesting observed features relatively 
to a H-He NLTE model with $T_{eff}=38000K, \log g=5.3, A(He)=N(He/H)=40$.
This model is a certain preliminary compromise to reproduce different 
spectral lines.
The following problems arise if one tries to fit our observations
with pure H-He NLTE model.
HeII 4686 and HeII 4541 lines are very strong while some HeI lines like
HeI 3867-71,3926 are almost absent (see Fig.\ref{f1}) and one needs 
temperatures above 40000K to improve the fit significantly.
On the other hand, red part of our spectra contains strong HeI
but weak HeII 4338 line of Pickering series and weak HeII 6527
(see Figs. \ref{f2}, \ref{f3}, \ref{f4}) 
what strongly favors temperatures below 36000K.
Strong HeI lines have generally narrower wings and would prefer 
$\log g \approx 5.$ while $H\gamma$ is broader and best reproduced with 
$\log g \approx 5.5$. Also it seems that our model predict generally weaker
Balmer and HeI lines with increasing wavelength.
$H\alpha$ exhibits a central emission reversal (see Fig.\ref{f4}),
which is a well-know NLTE effects found at hot white dwarfs and
subdwarfs (e.g. Lanz \& Hubeny 1995; Lanz et al. 1997). 
The absorption profile of $H\alpha$ is asymmetric, with the blue
wing being much deeper. This could be a signature of a stellar wind.
Asymmetry in the HeI 4471 and 4388 seems to be qualitatively reproduced
by the atomic line broadening data.
NIII/NII ionization balance speaks in favor of higher temperatures.
Nitrogen is very abundant and would certainly be important opacity source.

Perhaps, an atmosphere model with much steeper temperature gradient
due to, for instance, line blanketing could enhance HeII 4686 and HeI 4471
lines simultaneously
and suppress some temperature inconsistencies. Also weak Pickering line
HeII 4338 might be understood as it forms on the background of hydrogen 
opacity and consequently would originate from higher and cooler layers.
Blue-red discrepancies in Balmer and some HeI lines lines could be caused 
bye wrong correction on the stimulated emission resulting from the 
inappropriate model of the atmosphere as mentioned by Lanz et al. (1997).
More sophisticated line blanketed model atmospheres are called for.

\begin{figure}
\plotone{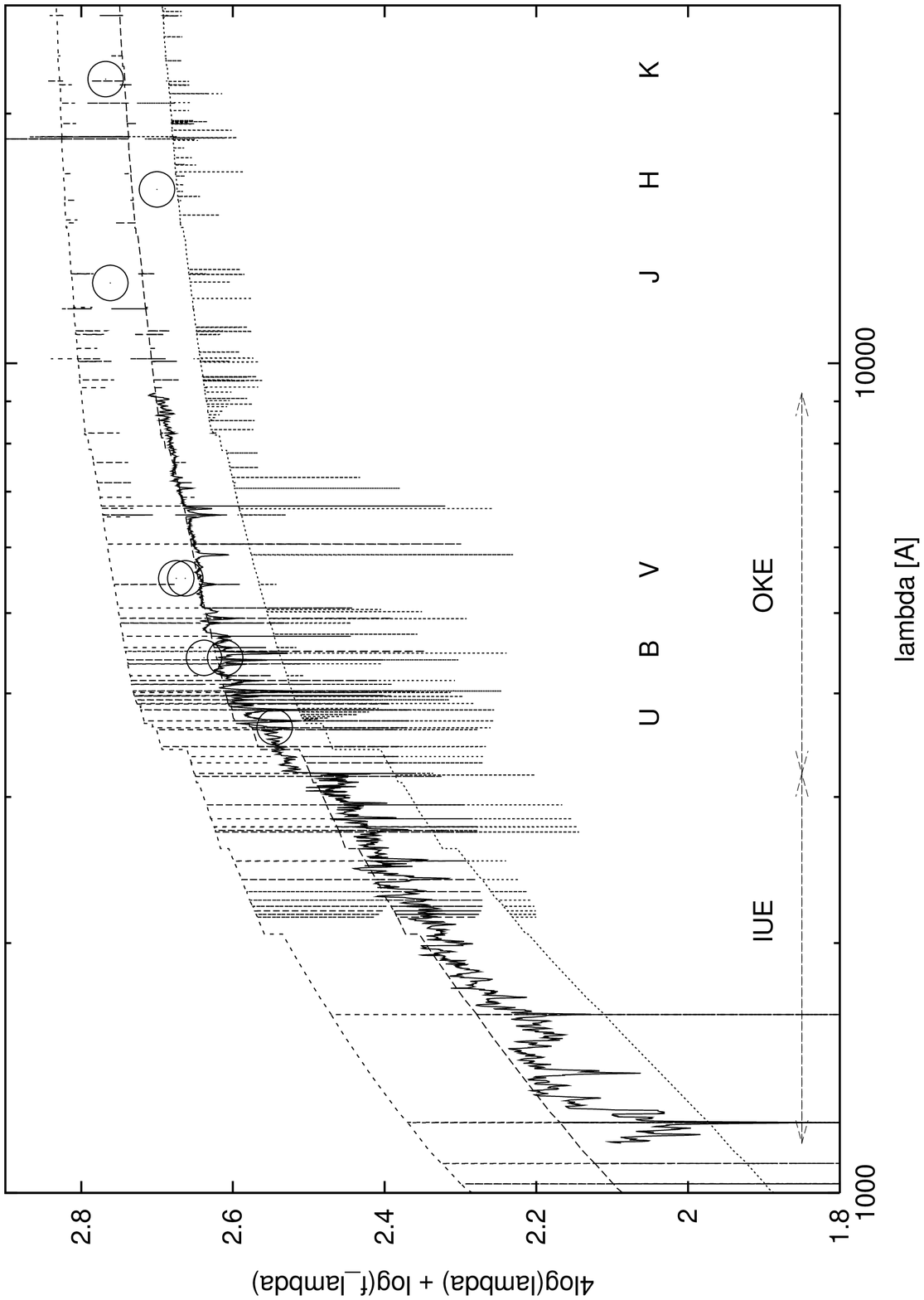}
\caption{Observations versus absolute 
fluxes from three model atmospheres from the top: 
$T_{eff}=43000,38000,34000K; \log g=6.7, 5.3, 5.5; N(He/H)=10,40,40$. 
Fluxes are in $erg/cm^{2}/s/\AA$ and are multiplied by $\lambda^{4}[\AA^{4}]$. 
Model fluxes are sifted bye 21.04 dex.}
\label{f5}
\end{figure}

Another, solution of some of the mentioned inconsistencies could be 
proposed. The spectrum could be a composite of two different spectra
one having the temperature in excess of 40000K the other below about 36000K.
This seems to be in accordance with Ulla \& Thejll (1998) who suggested 
that this star is a binary candidate based on its JHK excess.
On the other hand, their excess was found comparing the observed data
with Kurucz line blanketed models with normal He abundance.
We collected low resolution IUE, ground based spectrophotometric and 
UBVJHK photometric data from the sources mentioned  in the Sec.\ref{s1} 
and compare it with our model atmosphere in Fig.\ref{f5}.  
UBV data were calibrated using an average A0V star as a comparison and 
JHK filters 
were calibrated on Vega. Comparison object absolute fluxes were taken from 
Cox (2000). Table \ref{t3} lists the calibration constants derived
to get $I_{\lambda}$ $[erg\, cm^{-2}\,s^{-1}\,\AA^{-1}]$ from $m_{\lambda}$
-- magnitude in the particular filter using the following equation:
\begin{equation}
\log I_{\lambda}=-0.4m_{\lambda}-q
\end{equation}
Although there might be a J excess of about $0.1^{m}$ it seems rather 
questionable as it is almost within the error comparable to the possible 
photometric variability of this star and precision of such calibrations 
indicated in the figure by the radius of the open circles. 
Radial velocities (see Table \ref{t1}) of our spectra are constant 
within the precision of our measurements and do not exhibit potential 
orbital motion.
Moreover, this hypothesis would not, probably, be able to account fully
for some strong observed HeI lines not would it help to get rid fully
of strong synthetic HeII 4338... and could only partially improve
the situation.
Fig. \ref{f5} also reveals that effective temperature adopted for this star
is not bad although it could be slightly higher. Detailed
synthetic spectra in the IUE region would help to locate the continuum here.

\begin{table}
\caption{Photometry and calibration of absolute fluxes.
Calibration comparison standards used, their magnitudes and 
calibration constants derived are listed.}
\begin{center}
\begin{tabular}{llll}
\hline
filter  & calib& mag.& q \\
\hline
U       &  A0V & -0.04 & 8.391 \\
B       &  A0V & -0.02 & 8.202 \\
V       &  A0V &  0.00 & 8.426 \\
R       &  A0V & -0.02 & 8.765 \\
I       &  A0V &  0.00 & 9.076 \\
J       &  Vega & 0.02 & 9.472 \\
H       &  Vega & 0.02 & 9.931 \\
K       &  Vega & 0.02 & 10.375 \\
\hline
\end{tabular}
\end{center}
\label{t3}
\end{table}

Similar effects of a two component model atmosphere could be expected 
if there is a convection resulting
in some kind of solar like granulation with hotter granules and cooler
inter-granules. However, Groth et al. (1985) found that convection
- when present - is a very ineffective energy transport mechanism in the
atmospheres of such type of stars.

From absolute fluxes and parallax we can determine stellar parameters of 
the star as well. The slope of the Paschen continuum or B-V color index are
not very sensitive to the temperature of such hot stars because
they are in the Rayleigh-Jeans region of the Planck function and 
$F_{\lambda} \sim T_{eff}$. Absolute flux observed on the Earth is
$f_{\lambda}=(R/D)^{2}F_{\lambda}$ and one could determine reliable 
radius R of the star if he knows the distance D and compares absolute 
and theoretical fluxes $F_{\lambda}$.  As 
\begin{equation}
\frac{R}{R_{\odot}}=4.433 \times 10^{7} \sqrt{\frac{f_{\lambda}}{F_{\lambda}}}
\frac{1}{\pi^{"}}
\end{equation}
\begin{equation}
\frac{d R}{R}=\frac {d f_{\lambda}}{2f_{\lambda}} - 
\frac{dT_{eff}}{2T_{eff}} - \frac{d\pi^{"}}{\pi^{"}}
\end{equation}
generally a 10\% error in $T_{eff}$ results in about 5\% error 
in the stellar radius and parallax puts most severe constraints on the
precision of this method (Muthsam \& Weiss 1978).
We tried to fit the observed absolute data in Fig.\ref{f5} with our
preliminary model and found 
$f_{\lambda}/F_{\lambda}=10^{-21.04 \pm 0.1}$. Assuming $D=111$ pc 
it results in $R=0.15 (\pm 0.04) R_{\odot}$.
Now, assuming $\log g=5.3$ we get an extremely low mass, 
\begin{equation}
\frac{M}{M_{\odot}}=\frac{g}{2.74 \times 10^{4}}\frac{R^{2}}{R_{\odot}^{2}}
=0.16.
\end{equation}
This mass is lower than the lower limit for the core He burning,
$M=0.3M_{\odot}$ (Heber 1992).
Nevertheless it is
easily possible that the gravity is
higher(lower) by a factor of 2-3 due to above mentioned uncertainties 
in determining $\log g$ from H and He line profiles.

The actual luminosity of the star is:
\begin{equation}
L=4 \pi R^{2} \sigma T_{e}^{4}=4.2 \times 10^{1} L_{\odot}.
\end{equation}
Consequently, absolute bolometric and visual magnitudes of the star are:
\begin{equation}
M_{bol}=-2.5 \log L/L_{\odot}+ 4.74=+0.7^{m} 
\end{equation}
\begin{equation}
M_{V}=m_{V}-5 (\log D [pc]- 1)=+4.43^{m}
\end{equation}

Finally, high resolution spectra enabled us to set a rather low limit on 
the rotation of the star. Assuming zero microturbulence and Gaussian
instrumental profile with FWHM=0.3\AA~ we obtained $vsini=15 km/s$.
On the other hand microturbulence itself cannot be higher than the same
value.

\section{Conclusions}
We can conclude that H-He NLTE model cannot provide a satisfactory
description of the complex spectral features of this star.
Despite of this fact some parameters of the star and its atmosphere
such as its radius are rather insensitive to the uncertainty in 
the effective temperature and could be estimated. Another parameters,
for instance, gravity, seems quite
different from the value mentioned in the literature.

\acknowledgments

JB gratefully acknowledges the support of the IAU travel grant,
VEGA Grant No. 7107 and APVT-51-000802 project. 
Computational resources of the Computing Centre of 
the Slovak Academy of Sciences were partly used in some calculations.

\begin{references}
\reference Bartkevicius A., Lazauskaite R., 1996, Baltic Astronomy 5, 1
\reference Bohlin R.C., Harris A.W., Holm A.V., Gry C., 1990, ApJS 73, 413\\
(http://www.eso.org/observing/standards/spectra/)
\reference Bartolini C., Bonifazi A., Pecci F.F., Oculi L., Piccioni A., Serra R., 
Dantona F., 1982,  Astrophys. and Space Science 83, 287
\reference Colina L., Bohlin R.C., 1994, AJ 108, 1931
\reference Cox A.N., 2000, Allen's Astrophysical Quantities, Springer-Verlag New York,
Inc., p. 150, 387
\reference Diplas A., Savage B.D., 1994,  ApJS 93, 211
\reference Dworetsky M.M., Whitelock P.A., Carnochan D.J., MNRAS 201, 901
\reference Elkin E.G., 1998, Contrib. Astron. Obs. Skalnate Pleso, 27, 452
\reference ESA, 1997 in {\it The Hipparcos and Tycho Catalogues}, ESA SP-1200
\reference Galazutginov G.A., 1992, Preprint SAO 92 
\reference Greenstein J.L., 1960, in: Stellar atmospheres, ed. Greenstein J.L.,
Univ. Chicago press, Chapt. 19.
\reference Greenstein J.L., Sargent A.I., 1974, ApJS 28, 157
\reference Groth H.G., Kudritzki R.P., Heber U., 1985, AA 152,107
\reference Heber U., 1992, in: Atmospheres of Early type stars,
eds. U. Heber and C.S. Jeffery, Springer-Verlag Berlin Heidelberg, p.233
\reference Hubeny I., 1988, Comput. Phys. Comm., 52, 103
\reference Hubeny I., Lanz T., 1992, A\&A, 262, 501
\reference Hubeny I., Lanz T. 1995, ApJ, 439, 875
\reference Hubeny I., Lanz T., Jeffery C.S., 1995, Tlusty \& Synspec - A User's Guide
\reference Kaufmann J.P., Theil U., 1980, AASS 41, 271
\reference Kupka F., Piskunov N.E., Ryabchikova T.A., Stempels H.C., Weiss W.W., 
AAS 138, 119 (1999) (VALD-2) 
\reference Kurucz R.L., 1990, Trans. IAU, XXB, 168 (CD-ROM 23)
\reference Lanz T., Hubeny I., 1995, ApJ 439, 905
\reference Lanz T., Hubeny I., Heap S.R., 1997, ApJ 485, 843
\reference Muthsam H., Weiss W.W., 1978, in: Ap stars in the infrared, 
eds. W.W. Weiss, T.J. Kreidl, Univ. of Vienna, p. 37
\reference Oke J.B., 1990, AJ 99, 1621
\reference Peterson A.V., 1970, Thesis, California Institute of Technology
\reference Richter D., 1971, AA 14, 415
\reference Thejll P., Bauer F., Saffer R., Liebert J., Kunze D., Shipman H.L., 1994, 
ApJ 433, 819
\reference Tobin W., 1985, AASS 60, 459
\reference Ulla A., Theill P., 1998, AAS 132, 1
\reference Varosi F., Lanz T., Dekoter A., Hubeny I., Heap S., 1995, MODION (NASA
Goddard SFC)
\newline URL=ftp://idlastro.gsfc.nasa.gov/pub/contrib\\/varosi/modion/
\reference Vlasyuk V.V., 1993, Bulletin of SAO 36, p. 107
\end{references}
\end{document}